\documentclass{ws-m3as}
\usepackage{epstopdf}

\makeatletter
\renewcommand*\env@matrix[1][*\c@MaxMatrixCols c]{%
  \hskip -\arraycolsep
  \let\@ifnextchar\new@ifnextchar
  \array{#1}}
\makeatother

\begin{document}

\markboth{A. M. Blokhin, D. L. Tkachev}{Stability of a Poiseuille-type flow for an MHD model}

%
%

\title{Stability of a Poiseuille-type flow for a MHD model of an incompressible polymeric fluid }

\author{A. M. Blokhin}

\address{Sobolev Institute of Mathematics, 630090, 4 Acad. Koptuyg Avenue, Novosibirsk, 630090, Russia\\
Mechanics and Mathematics Department, Novosibirsk State University, 630090, 1 Pirogova Str., Novosibirsk, Russia\\
blokhin@math.nsc.ru}

\author{D. L. Tkachev}

\address{Sobolev Institute of Mathematics, 630090, 4 Acad. Koptuyg Avenue, Novosibirsk, 630090, Russia\\
Mechanics and Mathematics Department, Novosibirsk State University, 630090, 1 Pirogova Str., Novosibirsk, Russia\\
tkachev@math.nsc.ru}

\maketitle

\begin{history}
https://doi.org/10.1016/j/euromechflu.2019.12.006
© 2019. This manuscript version is made available under the CC-BY-NC-ND 4.0 license http://creativecommons.org/licenses/by-nc-nd/4.0/
\end{history}

\begin{abstract}
We study a generalization of the Pokrovski--Vinogradov model for flows of solutions and melts of an incompressible viscoelastic polymeric medium to nonisothermal flows in an infinite plane channel under the influence of magnetic field. For the linearized problem (when the basic solution is an analogue of the classical Poiseuille flow for a viscous fluid described by the Navier-Stokes equations) we find a formal asymptotic representation for the eigenvalues under the growth of their modulus. We obtain a necessary condition for the asymptotic stability of a Poiseuille-type shear flow.
\end{abstract}

\keywords{Incompressible viscoelastic polymeric fluid; rheological relation; magnetohydrodynamic flow.}

\ccode{Mathematics Subject Classification 2010: 35S99, 76J99, 76T20}

\section*{Introduction}	

In this work we study a generalization of the structurally phenomenological Pokrovski--Vinogradov model describing flows of melts and solutions of incompressible viscoelastic polymeric media to the nonisothermal case under the influence of magnetic field. In the Pokrovski--Vinogradov model, the polymeric medium is considered as a suspension of polymer macromolecules which move in an anisotropic fluid produced, for example, by a solvent and other macromolecules. The influence of environment on a real macromolecule is modeled by the action on a linear chain of Brownian particles each of which represents a large enough part of the macromolecule. Brownian particles often called ``beads'' are connected to each other by elastic forces called ``springs''. In the case of slow motions the macromolecule is modeled as a chain of two particles called ``dumbbell''.

The physical representation of linear polymer flows described above results in the formulation of the Pokrovski--Vinogradov rheological model \cite{3,1,2}:
\begin{equation}
\rho(\frac{\partial}{\partial t}v_{i} + v_{k}\frac{\partial}{\partial x_{k}}v_{i}) = \frac{\partial}{\partial x_{k}}\sigma_{ik}, \quad \frac{\partial v_{i}}{\partial x_{i}} = 0, \label{0.1}
\end{equation}
\begin{equation}
\sigma_{ik} = -p\delta_{ik} + 3 \frac{\eta_{0}}{\tau_{0}}a_{ik}, \label{0.2}
\end{equation}
\begin{equation}
\frac{d}{dt}a_{ik} - v_{ij}a_{jk} - v_{kj}a_{ji} + \frac{1 + (k - \beta)I}{\tau_{0}}a_{ik} = \frac{2}{3} \gamma_{ik} - \frac{3\beta}{\tau_{0}}a_{ij}a_{jk}, \label{0.3}
\end{equation}
where $\rho$ is the polymer density, $v_{i}$ is the $i$-th velocity component, $\sigma_{ik}$ is the stress tensor, $p$ is the hydrostatic pressure, $\eta_{0}$, $\tau_{0}$ are the initial values of the shear viscosity and the relaxation time respectively for the viscoelastic component, $v_{ij}$ is the tensor of the velocity gradient, $a_{ik}$ is the symmetrical tensor of additional stresses of second rank, $I = a_{11} + a_{22} + a_{33}$ is the first invariant of the tensor of additional stresses, $\gamma_{ik} = \frac{v_{ik} + v_{ki}}{2}$ is the symmetrized tensor of the velocity gradient, $k$ and $\beta$ are the phenomenological parameters taking into account the shape and the size of the coiled molecule in the dynamics equations of the polymer macromolecule.

Structurally, the model consists of the incompressibility and motion equations \eqref{0.1} as well as the rheological relations \eqref{0.2}, \eqref{0.3} connecting kinematic characteristics of the flow and its inner thermodynamic parameters.

Some generalizations of model \eqref{0.1} -- \eqref{0.3} provide good results in numerical simulations  for viscosymetric flows \cite{4}. For example, such a generalization is a model for which in equation \eqref{0.2} we add a term taking into account the so-called shear viscosity and the parameter  $\beta$ is additionally dependent on the first invariant of the anisotropy tensor. Therefore,
we may believe  that modifications of the basic Polrovski--Vinogradov model could be useful for modeling the polymer motion in complex deformation conditions, e.g., for stationary and non-stationary flows in circular channels, flows in channels with a fast change of sectional area and flows with a free boundary. An important feature of such flows is their two- and three dimensional character.

In this work, we consider one of such generalizations that takes into account the influence of the heat and the magnetic field on the  polymeric fluid motion (see Sect. 1 for more details). Our main interest is an analogue of the Poiseuille flow which is the well-known shear flow in an infinite channel. It turns out that in our case this flow has a number of features. For example, computations show that for some values of parameters the velocity profile is stretched in the direction opposite to the forces of pressure (see Sect. 1).

Our main results are given in Sect. 2. Firstly, we get an asymptotic representation of the spectrum of the  problem linearized about the the chosen basic solution which is the Poiseuille-type flow. Secondly, as the result we get a condition whose fulfilment guarantees that the basic solution is asymptotically stable by Lyapunov in the chosen class of perturbations periodic with respect the variable changing along the infinite plane channel. The last section is devoted to the proof of the theorems formulated in Sect. 2.

Overall, our work continues the study of Lyapunov's stability of shear flows for both the original Pokrovski--Vinogradov model and its various generalizations described in \cite{11,8,10,7,6,9}.

It should be noted that for the case of viscous fluid there is the known Krylov's result \cite{12}  about the linear Lyapunov's instability  of the Poiseuille flow for large enough Reynolds numbers confirming Heisenberg's hypothesis \cite{13} (a refinement of this result was obtained in \cite{14}).

\section{Nonlinear model of the polymeric fluid flow in a plane channel under the presence of an external magnetic field}

Using the results from \cite{23,17,16,19,18} and \cite{20}, let us formulate a mathematical model describing magnetohydrodynamic flows of an incompressible polymeric fluid for which, as in \cite{21}, in the equation for the inner energy we introduce some dissipative terms. In a dimensionless form this model reads (we keep the notations from \cite{22}):
\begin{equation}
div\vec{\boldsymbol{u}} = u_{x} + v_{y} = 0, \label{1.1}
\end{equation}
\begin{equation}
div\vec{H} = L_{x} + M_{y} = 0, \label{1.2}
\end{equation}
\begin{equation}
\frac{d\vec{\boldsymbol{u}}}{dt} + \nabla P = div(Z\Pi) + \sigma_{m}(\vec{H}, \nabla)\vec{H} + Gr(Z-1)\begin{pmatrix}0\\1\end{pmatrix}, \label{1.3}
\end{equation}
\begin{equation}
\frac{da_{11}}{dt} - 2A_{1}u_{x} - 2a_{12}u_{y} + L_{11} = 0, \label{1.4}
\end{equation}
\begin{equation}
\frac{da_{22}}{dt} - 2A_{2}v_{y} - 2a_{12}v_{x} + L_{22} = 0, \label{1.5}
\end{equation}
\begin{equation}
\frac{da_{12}}{dt} - A_{1}v_{x} - A_{2}u_{y} + \frac{\widetilde{K}_{I}a_{12}}{\bar{\tau}_{0}(Z)} = 0, \label{1.6}
\end{equation}
\begin{equation}
\frac{dZ}{dt} = \frac{1}{Pr}\Delta_{x,y}Z + \frac{A_{r}}{Pr}ZD_{\Gamma} + \frac{A_{m}}{Pr}\sigma_{m}D_{m}, \label{1.7}
\end{equation}
\begin{equation}
\frac{d\vec{H}}{dt} - (\vec{H},\nabla)\vec{\boldsymbol{u}} - b_{m}\Delta_{x,y}\vec{H} = 0. \label{1.8}
\end{equation}
where $t$ is the time, $u$, $v$ and $L$, $1+M$ are the components of the velocity vector $\vec{\boldsymbol{u}}$ and the magnetic field $\vec{H}$ respectively in the Cartesian coordinate system $x,y$;
$$P = \rho + \sigma_{m}\frac{L^{2} + (1+ M)^{2}}{2},$$ 
$\rho$ is the pressure;
$a_{11}$, $a_{22}$, $a_{12}$ are the components of the symmetrical anisotropy tensor of second rank;
$$\Pi = \frac{1}{Re}(a_{ij}),\quad i,j = 1,2;\quad L_{ii} = \frac{K_{I}a_{ii} + \beta(a_{ii}^{2} + a_{12}^{2})}{\bar{\tau}_{0}(Z)},\quad i = 1,2;$$
$$K_{I} = W^{-1} + \frac{\bar{k}}{3}I,\quad \bar{k} = k - \beta ;$$ $I = a_{11} + a_{22}$ is the first invariant of the anisotropy tensor; $k$, $\beta$ ($0 < \beta < 1$) are the phenomenological parameters of the rheological model (see \cite{1});
$A_{i} = W^{-1} + a_{ii}$, $i = 1,2$; $Z = \frac{T}{T_{0}}$; $T$ is the temperature, $T_{0}$ is an average temperature (room temperature; we will further assume that $T_{0} = 300$ K);\\
$\widetilde{K}_{I} = K_{I} + \beta I$; $\bar{\tau}_{0}(Z) = \frac{1}{ZJ(Z)}$, $J(Z) = \exp\{\bar{E}_{A}\frac{Z-1}{Z}\}$,\\
$\bar{E}_{A} = \frac{E_{A}}{T_{0}}$, $E_{A}$ is the activation energy;\\
$Re = \frac{\rho u_{H}l}{\eta_{0}^{*}}$ is the Reynolds number;\\
$W = \frac{\tau_{0}^{*}u_{H}}{l}$ is the Weissenberg number;\\
$Gr = \frac{Ra}{Pr}$ is the Grasshoff number;\\
$Pr = \frac{lu_{H}\rho c_{v}}{\varepsilon} = \frac{c_{v}\eta_{0}^{*}Re}{\varepsilon}$ is the Prandtl number;\\
$Ra = \frac{lbgT_{0}Pr}{u_{H}^{2}}$ is the Rayleigh number;\\
$A_{r} = \frac{\alpha u_{H}^{2}Pr}{ReT_{0}c_{v}} = \frac{\alpha u_{H}^{2}\eta_{0}^{*}}{T_{0}\varepsilon}$, $A_{m} = \frac{\alpha_{m} u_{H}^{2}Pr}{T_{0}c_{v}}$;\\
$D_{\Gamma} = a_{11}u_{x} + (v_{x} + u_{y})a_{12} + a_{22}v_{y}$;\\
$D_{m} = L^{2}u_{x} + L(1 + M)(v_{x} + u_{y}) + (1 + M)^{2}v_{y}$;\\
$\rho(= const)$ is the medium density;\\
$\varepsilon$ is the coefficient of thermal conductivity  of the polymeric fluid;\\
$b$ is the coefficient thermal expansion of the polymeric fluid;\\
$g$ is the gravity constant;\\
$\eta_{0}^{*}$, $\tau_{0}^{*}$ are the initial values for the shear viscosity and the relaxation time for the room temperature $T_{0}$ (see \cite{1,2});\\
$l$ is the characteristic length, $u_{H}$ is the characteristic velocity;\\
$\sigma_{m} = \frac{\mu\mu_{0}H_{0}^{2}}{\rho u_{H}^{2}}$ is the magnetic pressure coefficient;\\
$b_{m} = \frac{1}{Re_{m}}$, $Re_{m} = \sigma_{m}\mu\mu_{0}u_{H}l$ is the magnetic Reynolds number;\\
$\mu_{0}$ is the magnetic penetration in vacuum;\\
$\mu$ is the magnetic penetration of the polymeric fluid;\\
$\sigma$ is the electrical conductivity of the medium;\\
$\alpha$ is the thermal equivalent of work (see \cite{7});\\
$\alpha_{m}$ is the magnetothermal equivalent of work;\\
$c_{v}$ is the heat capacity;\\
$\frac{d}{dt} = \frac{\partial}{\partial t} + (\vec{\boldsymbol{u}}, \nabla) = \frac{\partial}{\partial t} + u\frac{\partial}{\partial x} + v\frac{\partial}{\partial y}$,\\
$\Delta_{x,y} = \frac{\partial^{2}}{\partial x^{2}} + \frac{\partial^{2}}{\partial y^{2}}$ is the Laplace operator.\\
The variables $t$, $x$, $y$, $u$, $v$, $p$, $a_{11}$, $a_{22}$, $a_{12}$, $L$, $M$ in system \eqref{1.1}--\eqref{1.8} correspond to the following values: $\frac{l}{u_{H}}$, $l$, $u_{H}$, $\rho u_{H}^{2}$, $\frac{W}{3}$, $H_{0}$, where $H_{0}$ is the characteristic magnitude of the magnetic field  (see fig. 1).

\begin{figure}[h]
\setlength{\unitlength}{0.1\textwidth}
\centerline{\includegraphics[width=0.5\textwidth]{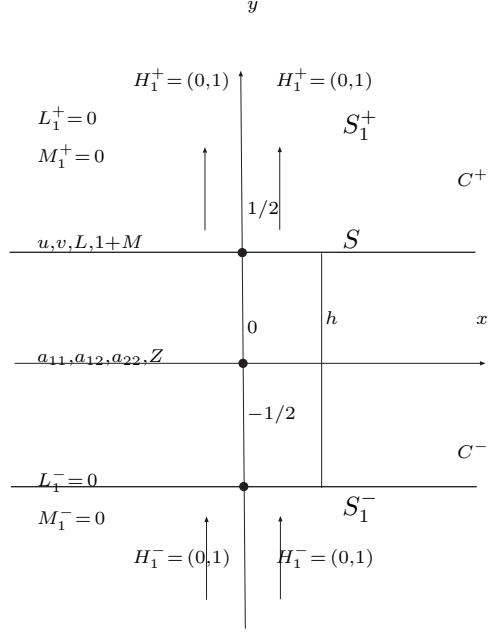}
\put(-0.3, 1.8){$\scriptstyle C^{-}$}
\put(-0.1, 3.2){$\scriptstyle x$}
\put(-0.3, 4.65){$\scriptstyle C^{+}$}
\put(-1.5, 1.2){$S_{1}^{-}$}
\put(-1.68, 3.2){$\scriptstyle h$}
\put(-1.5, 4){$S$}
\put(-1.5, 5.2){$S_{1}^{+}$}
\put(-2.5, 2.2){$\scriptstyle-1/2$}
\put(-2.5, 3.1){$\scriptstyle0$}
\put(-2.5, 4.35){$\scriptstyle1/2$}
\put(-2.5, 6.5){$\scriptstyle y$}
\put(-2.2, 5.7){$\scriptstyle H_{1}^{+} =\, (0,1)$}
\put(-2.2, 0.7){$\scriptstyle H_{1}^{-} =\, (0,1)$}
\put(-3.7, 0.7){$\scriptstyle H_{1}^{-} =\, (0,1)$}
\put(-3.7, 5.7){$\scriptstyle H_{1}^{+} =\, (0,1)$}
\put(-4.7, 1.1){$\scriptstyle M_{1}^{-} =\, 0$}
\put(-4.7, 1.5){$\scriptstyle L_{1}^{-} =\, 0$}
\put(-4.7, 2.8){$\scriptstyle a_{11}, a_{12}, a_{22}, Z$}
\put(-4.7, 4){$\scriptstyle u, v, L, 1 + M$}
\put(-4.7, 4.9){$\scriptstyle M_{1}^{+} =\, 0$}
\put(-4.7, 5.3){$\scriptstyle L_{1}^{+} =\, 0$}
} 
\caption{Plane channel}
\end{figure}

\begin{remark}
The magnetohydrodynamic equations \eqref{1.1} -- \eqref{1.8} are derived with the use of the Maxwell equations (see \cite{16,18}). The magnetic induction vector $\vec{B}$ is represented as
\begin{equation}
\vec{B} = \mu\mu_{0}\vec{H} = (1 + \chi)\mu_{0}\vec{H}, \label{1.9}
\end{equation}
where $\chi$ is the magnetic susceptibility, and (see \cite{23,24}) $\chi = \frac{\chi_{0}}{Z}$, $\chi_{0}$ is the magnetic susceptibility for the room temperature $T_{0}$(= 300 K). We will further assume that for the polymeric fluid $\mu =1$ ($\chi_{0} = 0$).
\end{remark}

\begin{remark}
Our main problem is the problem of finding solutions to the mathematical model \eqref{1.1}--\eqref{1.8} describing magnetohydrodynamic flows of an incompressible polymeric fluid in a plane channel with the depth $1(l)$ and bounded by the horizontal walls which are the electrodes $C^{+}$ and $C^{-}$ along which we have  electric currents with the current strength  $J^{+}$ and $J^{-}$ respectively (see Fig. 1).

External with respect to the chanel $S$ areas $S_{1}^{+}$, $S_{1}^{-}$ are under the influence of the magnetic fields with components $L_{1}^{+} = 0$, $M_{1}^{+}$, and $M_{1}^{+}|_{y = \frac{1}{2}+0} = -1 + \frac{1 + M(\frac{1}{2})}{1 + \chi_{0}^{+}}$, $L_{1}^{-} = 0$, $M_{1}^{-}$ and $M_{1}^{-}|_{y=-\frac{1}{2}+0} = -1 + \frac{1 + M(-\frac{1}{2})}{1 + \frac{\chi_{0}^{-}}{1 + \bar{\theta}}}$. Values of temperature $Z$ on the sides of the chanel will be defined below with boundary conditions \eqref{1.10}. Acquired correlations  between boundary values of $M_{1}^{+} + 1$, $1 + M(\frac{1}{2})$, and $M_{1}^{-} + 1$, $1 + M(-\frac{1}{2})$ correspondingly arise due to the equality \eqref{1.9} and continuity of the normal component for the magnetic induction vector on the sides of the chanel.

The domains $S_{1}^{+}$ and $S_{1}^{-}$ external to the channel  are magnets with the magnetic susceptibilities $\chi_{1}^{+}$ and $\chi_{1}^{-}$. On the walls of the channel the following boundary conditions hold:
\begin{equation}
\label{1.10}
\left\{\begin{array}{l}
y = \pm\frac{1}{2}: \quad \vec{u} = 0 \quad (\mbox{no-slip condition}),\\
y = \frac{1}{2}: \qquad Z = 1 \quad (T = T_{0}),\\
y = -\frac{1}{2}: \quad Z = 1 + \bar{\theta} \quad (\bar{\theta} = \frac{\theta}{T_{0}}, \, \theta = T - T_{0}).
\end{array}\right.
\end{equation}
We have the temperature $T = T_{0}$ in the domain $S_{1}^{+}$ where as on the electrode $C^{+}$ we have:
$$
y = -\frac{1}{2}: Z = 1 + \bar{\theta}, \bar{\theta} = \frac{\theta}{T_{0}}, \theta = T - T_{0},
$$
i.e., for $\bar{\theta} > 0$ there is  heating from below ($T$ is the temperature in the domain $S_{1}^{-}$ and on the electrode $C^{-}$), and for $\bar{\theta} < 0$ there is heating from above.
\end{remark}

\begin{remark}
We will consider the electrodes $C^{+}$ and $C^{-}$ as the boundaries between two uniform isotropic magnetics. Therefore, on the boundaries $C^{+}$ and $C^{-}$ the following known conditions hold (see \cite{23, 25}):
\begin{equation}
\label{1.11}
\left\{\begin{array}{l}
y = \frac{1}{2}(C^{+}): \quad L = -J^{+}, \quad M_{y} = 0,\\
y = -\frac{1}{2}(C^{-}): \quad L = -J^{-}, \quad M_{y} = 0.
\end{array}\right.
\end{equation}
\end{remark}
We get the boundary condition $M_{y} = 0$ at $y = \pm\frac{1}{2}$ by assuming that relation \eqref{1.2} holds for $y = \pm\frac{1}{2}$ and by taking into account the conditions $L = -J^{+}$ $(y = \frac{1}{2})$ and $L = -J^{-}$ $(y = -\frac{1}{2})$ (see \eqref{1.11}) that gives us $M_{y} = 0$ ($y = \pm\frac{1}{2}$).
\begin{remark}
Let us show that
$$
\left\{\begin{array}{l}
d = L_{x} + M_{y} = 0 \quad \mbox{for } y = \pm\frac{1}{2},\\
d = 0 \quad \mbox{for } t=0, |y| < \frac{1}{2}, x\in R^{1};\\
d \to 0 \quad \mbox{for } |x|\to\infty, t > 0, |y| < \frac{1}{2}, x\in R^{1},
\end{array}\right.
$$
i.e., relation \eqref{1.2} follows from equations \eqref{1.1}, \eqref{1.8}.
\end{remark}
To prove this we apply the operator div to equation \eqref{1.8}. Taking into account \eqref{1.1}, we get
$$
d_{t} + (\vec{u},\nabla)d - b_{m}\Delta_{x,y}d = 0.
$$
Consequently,
$$
(d^{2})_{t} + div(d^{2}\vec{u} - 2b_{m}d\cdot\nabla d) + 2b_{m}|\nabla d|^{2} = 0.
$$
Integrating this relation with respect to $x$ from $-\infty$ to $+\infty$ and with respect to $y$ from $-\frac{1}{2}$ to $\frac{1}{2}$ gives $$
\frac{d}{dt}\left\{\int_{-\frac{1}{2}}^{\frac{1}{2}}\int_{-\infty}^{+\infty}d^{2}(t,x,y)dxdy\right\} + 2b_{m}\int_{-\frac{1}{2}}^{\frac{1}{2}}\int_{-\infty}^{+\infty}|\nabla d(t,x,y)|^{2}dxdy = 0.
$$
This implies
$$
\int_{-\frac{1}{2}}^{\frac{1}{2}}\int_{-\infty}^{+\infty}d^{2}(t,x,y)dxdy \leq 0,
$$
i.e., $d = 0$ for $t > 0$, $|y| < \frac{1}{2}$, $x\in R^{1}$.

If we consider $d$ as a function from a wider class of functions bounded on each infinite set $\{(t,x,y)| 0\leq t\leq T, -\infty < x < \infty, -\frac{1}{2}\leq y \leq \frac{1}{2}$\} (the parameter $T$ is being varied), then the fact that $d$ vanishes follows from the maximum principle for the heat equation
$$
d_{t} - b_{m}\Delta_{x,y}d = 0.
$$
This equation can be obtained by rewriting  relation \eqref{1.8} in the form
$$
\frac{d\vec{H}}{dt} - rot(\vec{u}\times\vec{H})-b_{m}\Delta_{x,y}\vec{H} = 0
$$
and using the operator div.

Stationary solutions of the mathematical model \eqref{1.1} -- \eqref{1.8} were studied in \cite{22}. Particular solutions (analogous to Poiseuille and Couette solutions for the Navier--Stoks equations system) were constructed there in the following form:
\begin{equation}
\label{1.12}
\left\{\begin{array}{l}
\vec{U}(t,x,y) = \hat{\vec{U}}(y),\\
p(t,x,y) = \hat{P}(y) + \hat{p}_{0} - \hat{A}x,
\end{array}\right.
\end{equation}
where\\
$\vec{U} = (u, v, a_{11}, a_{12}, a_{22}, Z, L, M)^{T}$,\\
$\hat{\vec{U}}(y) = (\hat{u}(y), \hat{v}(y), \hat{a}_{11}(y), \hat{a}_{12}(y), \hat{a}_{22}(y), \hat{Z}(y), \hat{L}(y), \hat{M}(y))^{T}$,\\
$\hat{P}(y)$ ($\hat{P}(0) = 0$) is some function that we will define further, $\hat{p}_{0}$ is the pressure on the channel axis for $y = 0$, $x = 0$ and $\hat{A} (>0)$ is the dimensionless constant drop of pressure on the segment $h$.

From \eqref{1.1} -- \eqref{1.8}, \eqref{1.10}, \eqref{1.11} we have the following relations for getting functions $\hat{u}(y)$, $\hat{a}_{11}(y)$, $\hat{a}_{12}(y)$, $\hat{a}_{22}(y)$, $\hat{Z}(y)$, $\hat{L}(y)$, $\hat{M}(y)$, $\hat{P}(y)$:
\begin{equation}
\label{1.13}
\begin{aligned}
&\frac{d}{dy}(\hat{Z}\hat{a}_{12} + (1 + \hat{\lambda}\sigma_{m}Re\hat{L}) = (\hat{Z}\hat{a}_{12} + (1 + \hat{\lambda}\sigma_{m}Re\hat{L})^{'} = -\hat{D},\\
&(\hat{P} - \frac{\hat{Z}\hat{a}_{22}}{Re} + \sigma_{m}\frac{\hat{L}^{2}}{2})^{'} = Gr(\hat{Z} -1), \quad \hat{P}(0) = 0,\\
&\hat{u}^{'} = \frac{\tilde{K}_{\hat{I}}J(\hat{Z})\hat{Z}\hat{a}_{12}}{\hat{A}_{2}}, \quad \hat{u}(\pm\frac{1}{2}) = 0,\\
&K_{\hat{I}}\hat{a}_{22} + \beta(\hat{g} + \hat{a}_{22}^{2}) = 0,\\
&K_{\hat{I}}\hat{a}_{11} + \beta(\hat{g} + \hat{a}_{11}^{2}) - 2\hat{g}\frac{\tilde{K}_{\hat{I}}}{\hat{A}_{2}} = 0,\\
&\hat{Z}^{''} + (a_{r}\hat{Z}\hat{a}_{12} + A_{m}\sigma_{m}(1 + \hat{\lambda})\hat{L})\hat{u}^{'} = 0, \, \hat{Z}(\frac{1}{2}) = 1, \, \hat{Z}(-\frac{1}{2}) = 1 + \bar{\theta},\\
&b_{m}\hat{L}^{''} + (1 + \hat{\lambda})\hat{u}^{'} = 0, \quad \hat{L}(\pm\frac{1}{2}) = -J^{\pm},\\
&\hat{M} = \hat{\lambda} = \chi_{0}^{+} = \frac{\chi_{0}^{-}}{1 + \bar{\theta}} = const.
\end{aligned}
\end{equation}
Here $\hat{D} = Re\cdot\hat{A}$, $K_{\hat{I}} = W^{-1} + \frac{\bar{k}}{3}\hat{I}$, $\hat{I} = \hat{a}_{11} + \hat{a}_{22}$,\\
$\tilde{K}_{\hat{I}} = K_{\hat{I}} + \beta\hat{I}$, $\hat{g} = \hat{a}_{12}^{2}$, $\hat{A}_{2} = W^{-1} + \hat{a}_{22}$,\\
$\chi_{0}^{+}$ and  $\chi_{0}^{-}$ are the magnetic susceptibilities for $T = T_{0}$ in the domains $S_{1}^{+}$ and $S_{1}^{-}$ respectively. A detailed analysis of relations \eqref{1.13} was performed in \cite{22}.

Some solutions or more precisely their components $\hat{u}$, $\hat{Z}$ and $\hat{L}$ are represented in Fig. 2, 3, 4, 5. Moreover, in the first (main) case (if we use the terminology from \cite{22}) $\hat{A} = 1$, $\hat{\lambda} = 1$, $\sigma_{m} = 1$, $Re = 1$, $W = 1$, $\beta = 0.5$, $A_{r} = 1$, $A_{m} = 1$, $\bar{\theta} = 1$, $b_{m} = 1$, $\bar{E}_{A} = 1$, $J^{+} = 2$, $J^{-} = 1$, and in the second, third and forth cases we change one of the parameters $\hat{A}$, $J^{+}$ and $\theta$ respectively by leaving other parameters unchanged.

\begin{figure}[h]
\centerline{\includegraphics[width=0.33\textwidth]{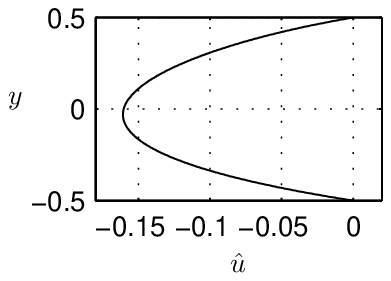}\includegraphics[width=0.33\textwidth]{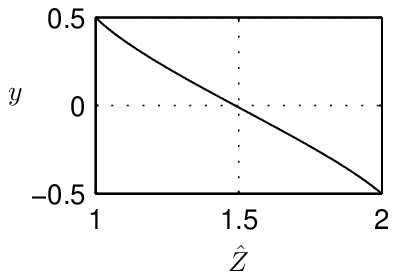}\includegraphics[width=0.33\textwidth]{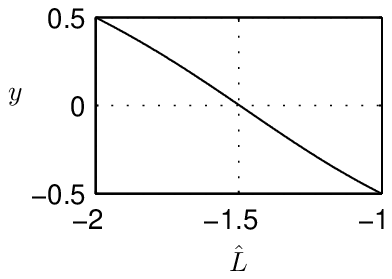}} 
\caption{Main case.}
\end{figure}

\begin{figure}[h]
\centerline{\includegraphics[width=0.33\textwidth]{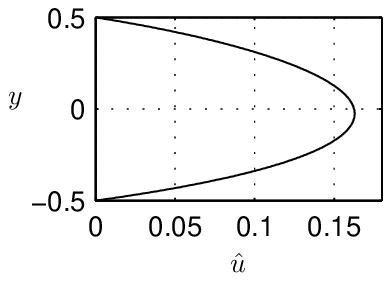}\includegraphics[width=0.33\textwidth]{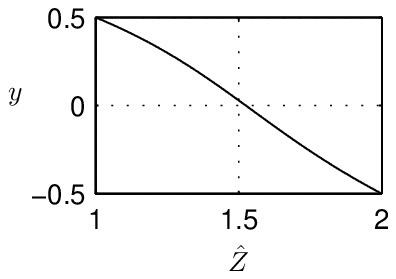}\includegraphics[width=0.33\textwidth]{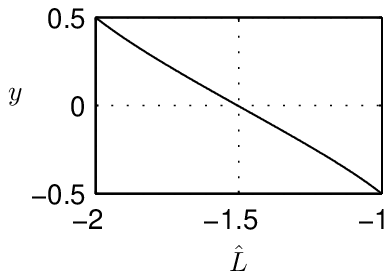}} 
\caption{Solution for $\hat{A}=3$.}
\end{figure}

\begin{figure}[h]
\centerline{\includegraphics[width=0.33\textwidth]{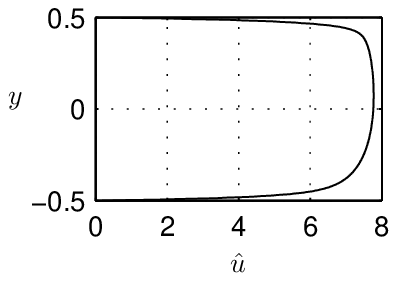}\includegraphics[width=0.33\textwidth]{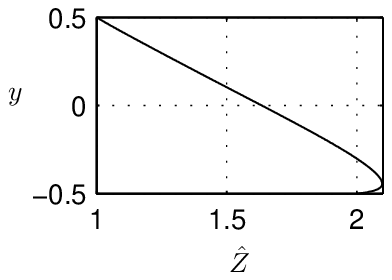}\includegraphics[width=0.33\textwidth]{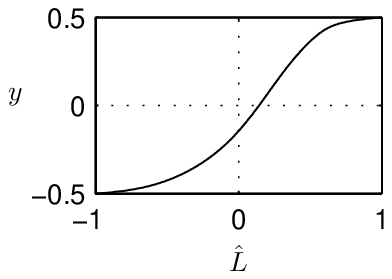}} 
\caption{Solution for $J^{+}=-1$.}
\end{figure}

\begin{figure}[h]
\centerline{\includegraphics[width=0.33\textwidth]{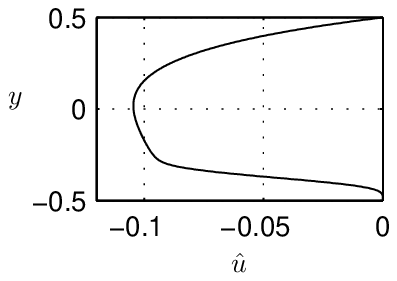}\includegraphics[width=0.33\textwidth]{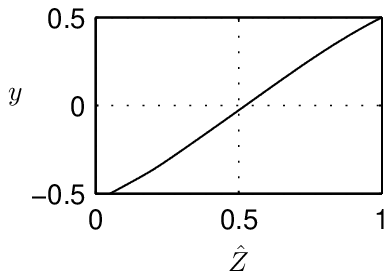}\includegraphics[width=0.33\textwidth]{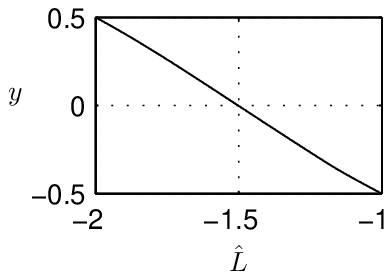}} 
\caption{Solution for $\bar{\theta}=-0.95$.}
\end{figure}

Let us list most important features of the behavior of stationary polymeric fluid flows. In \cite{20} the influence of  the parameter $\bar{E}_{A} = \frac{E_{A}}{T_{0}}$ (connected with the activation energy  $E_{A}$) on the form of the velocity profile was found. Unlike  \cite{20}, in our case the velocity profile loses symmetry which is a feature of the parabolic velocity profile of the Poiseuille flow for viscous fluid \cite{17}. This means that the solutions of \eqref{1.1} -- \eqref{1.8}, \eqref{1.10}, \eqref{1.11} have a wider range of interesting properties.

In the main case (see Fig. 2) the velocity profile is elongated in the direction opposite to that of pressure forces (due to the influence of magnetic field!). In Fig. 3 the pressure drop (its absolute value) is enlarged. This implies that the velocity profile is turned to the right. In Fig. 4 the absolute value of pressure is again small. But since the current's direction on the top electrode is now opposite to the previous one, the velocity profile is again turned to the right. At last, strong cooling of the bottom channel's boundary implies that the fluid velocity in the bottom part of the channel becomes close to zero (see Fig. 5).

We will further study Lyapunov's stability with respect to small perturbations of the stationary flow described above. Denoting small perturbations of all values by the same symbols, after linearization we obtain the following linear problem:
\begin{equation}\label{1.14}
\left\{\begin{aligned}
u_{t} &+ \hat{u}u_{x} - \hat{Z}(\alpha_{11})_{x} + \hat{Z}(\alpha_{22})_{x} - \hat{Z}(\alpha_{12})_{y} + \hat{u}^{'}v - \hat{Z}^{'}\alpha_{12} + \Gamma_{1} = 0,\\
v_{t} &+ \hat{u}v_{x} - \hat{Z}(\alpha_{12})_{x} + \Gamma_{2} = 0,\\
(\alpha_{11})_{t} &+ \hat{u}(\alpha_{11})_{x} - 2\hat{\alpha}_{1}u_{x} - 2\hat{\alpha}_{12}u_{y} + \hat{\alpha}_{11}^{'}v + R_{33}\alpha_{11} +\\
&+ R_{34}\alpha_{12} + R_{35}\alpha_{22} + r_{11}Z = 0,\\
(\alpha_{12})_{t} &+ \hat{u}(\alpha_{12})_{x} - \hat{\alpha}_{1}v_{x} - \hat{\alpha}_{2}u_{y} + \hat{\alpha}_{12}^{'}v + R_{43}\alpha_{11} +\\
&+ R_{54}\alpha_{12} + R_{45}\alpha_{22} + r_{12}Z = 0,\\
(\alpha_{22})_{t} &+ \hat{u}(\alpha_{22})_{x} - 2\hat{\alpha}_{12}v_{x} - 2\hat{\alpha}_{2}v_{y} + \hat{\alpha}_{22}^{'}v + R_{53}\alpha_{11} +\\
&+ R_{54}\alpha_{12} + R_{55}\alpha_{22} = 0,\\
Z_{t} &+ \hat{u}Z_{x} + \hat{Z}'v = \frac{1}{Pr}\Delta_{x,y}Z + \frac{Ar}{Pr}\hat{u}'\hat{a}_{12}Z + \frac{Ar}{Pr}\hat{Z}(\hat{a}_{11}u_{x} +\\
&+ \hat{a}_{12}v_{x} + \hat{a}_{12}u_{y} + \hat{a}_{22}v_{y} + \hat{u}'a_{12}) + \frac{Am}{Pr}\sigma_{m}(\hat{L}^{2}u_{x} +\\
&+ \hat{L}(1 + \hat{\lambda})v_{x} + \hat{L}(1 + \hat{\lambda})u_{y} + \hat{u}'((1 + \hat{\lambda})L + \hat{L}M) + (1 + \hat{\lambda})^{2}v_{y}), \\
L_{t} &+ \hat{u}L_{x} + v\hat{L}' - \hat{L}u_{x} - (1 + \hat{\lambda})u_{y} - \hat{u}'M - b_{m}\Delta_{x,y}L = 0,\\
M_{t} &+ \hat{u}M_{x} - \hat{L}v_{x} - (1 + \hat{\lambda})v_{y} - b_{m}\Delta_{x,y}M = 0, \\
u_{x} &+ v_{y} = 0, \quad t > 0, \quad |y| < \frac{1}{2}, \quad x \in R^{1};
\end{aligned}\right.
\end{equation}
\begin{equation}
\label{1.15}
u = v = Z = L = M_{y} = 0 \quad \mbox{при } y = \pm\frac{1}{2}, \, t > 0, x \in R^{1};
\end{equation}
where
$$
\begin{aligned}
\alpha_{ij} &= \frac{a_{ij}}{Re}, \hat{\alpha}_{ij} = \frac{\hat{a}_{ij}}{Re}, i,j = 1,2,\\
\hat{\alpha}_{i} &= \hat{\alpha}_{ii} + \kappa^{2}, \kappa^{2} = \frac{1}{WRe} i = 1,2,\\
\Gamma_{1} &= \Omega_{x}  - \hat{\alpha}_{11}Z_{x} - \hat{\alpha}_{12}Z_{y} - \hat{\alpha}_{12}'Z + \sigma_{m}(1 + \hat{\lambda})\omega_{m} - \sigma_{m}\hat{L}M,\\
\Gamma_{2} &= \Omega_{y} - \hat{\alpha}_{12}Z_{x} -\hat{\alpha}_{2}Z_{y} - (\hat{\alpha}_{12}^{'} + Gr)Z - \sigma_{m}\hat{L}\omega_{m} + \sigma_{m}\hat{L}'L,\\
\Omega &= P - \hat{Z}\alpha_{22}, \omega_{m} = M_{x} - L_{y}\\
R_{33} &= \hat{\bar{\chi}}_{0}(\hat{K}_{\hat{I}} + \hat{a}_{11}(\frac{\bar{k}}{3} + 2\beta)), R_{34} = -2\hat{u}' + 2\beta\hat{a}_{12}\chi_{0}^{*},\\
R_{35} &= \frac{\bar{k}}{3}\hat{a}_{11}\hat{\bar{\chi}}_{0}, \hat{\bar{\chi}}_{0} = \frac{1}{\bar{\tau}_{0}(\hat{Z})},\\
r_{11}& = (\hat{\bar{\chi}}_{0})'\frac{2\hat{\alpha}_{12}^{2}\tilde{K}_{\hat{I}}}{\hat{\alpha}_{2}}, \hat{\bar{\chi}}_{0}^{`} = \hat{\bar{\chi}}_{0}\frac{\bar{E}_{A} + \hat{Z}}{\hat{Z}^{2}}, \hat{\tilde{K}}_{\hat{I}} = \hat{K}_{\hat{I}} + \beta(\hat{a}_{11} + \hat{a}_{22}),\\
R_{43} &= \hat{a}_{12}\hat{\bar{\chi}}_{0}(\frac{\bar{k}}{3} + \beta), R_{45} = -\hat{u}' + \hat{a}_{12}\hat{\bar{\chi}}_{0}(\frac{\bar{k}}{3} + \beta), R_{44} = \hat{\bar{\chi}}_{0}^{'}\hat{\tilde{K}}_{\hat{I}},\\
r_{12} &= (\hat{\bar{\chi}}_{0})'\hat{\alpha}_{12}\hat{\tilde{K}}_{\hat{I}},
\end{aligned}
$$
\begin{equation}
\label{1.16}
R_{53} = \hat{\bar{\chi}}_{0}\hat{a}_{22}\frac{\bar{k}}{3}, R_{54} = 2\beta\hat{a}_{12}\hat{\bar{\chi}}_{0}, R_{55} = \hat{\bar{\chi}}_{0}(\hat{K}_{\hat{I}} + \hat{a}_{22}(\frac{\bar{k}}{3} + 2\beta)).
\end{equation}
\begin{remark}
The relation
$$
L_{x} + M_{y} = 0
$$
is not included into system \eqref{1.14} because due to the last three equations in \eqref{1.14} it holds for $t > 0$ if it was true for $t = 0$. That is, relation \eqref{1.2} is, in fact, a constraint on the initial data for $l$ and $M$.
\end{remark}
\begin{remark}
Unlike \cite{22}, the components $M_{1}^{+}$ and $M_{1}^{-}$ are not zero in the domains $S_{1}^{+}$ and $S_{1}^{-}$ (see Fig. 1):
\begin{equation}
\label{1.17}
\begin{aligned}
M_{1}^{+}\big|_{y = \frac{1}{2} +0} &= -1 + \frac{1 + M(\frac{1}{2})}{1 + \chi_{0}^{+}},\\
M_{1}^{-}\big|_{y = -\frac{1}{2} -0} &= -1 + \frac{1 + M(-\frac{1}{2})}{1 + \frac{\bar{\chi}_{0}^{-}}{(1 + \bar{\theta})}},
\end{aligned}
\end{equation}
where $M(\frac{1}{2})$ and $M(-\frac{1}{2})$ are the values of $M$ on the top and bottom electrodes respectively.
\end{remark}
\begin{remark}
Let us assume that the domains $S_{1}^{\pm}$ are filled with nonconducting mediums. Then, in view of the Maxwell equation \cite{25}, the small perturbations of  $M_{1}^{\pm}$ satisfy the Laplace equation and additional conditions at infinity:
\begin{equation}
\label{1.18}
\begin{aligned}
\Delta_{x,y}M_{1}^{+} &= 0 \, \mbox{in } S_{1}^{+}, \, M_{1}^{+}\to0 \, \mbox{for } y\to\infty,\\
\Delta_{x,y}M_{1}^{-} &= 0 \, \mbox{in } S_{1}^{-}, \, M_{1}^{-}\to0 \, \mbox{for } y\to-\infty.
\end{aligned}
\end{equation}
\end{remark}
If the components $M_{1}^{+}$ are periodic functions with respect to $x$, i.e.,
\begin{equation}
\label{1.19}
M_{1}^{\pm}(x,y) = \tilde{M}_{1}^{\pm}(y)e^{i\omega x}, \quad \omega \in R,
\end{equation}
then from relations \eqref{1.18} we get
$$
\begin{aligned}
\tilde{M}_{1}^{+}(y) &= M_{1}^{+}\big|_{y = \frac{1}{2} + 0}e^{-|\omega|(y - \frac{1}{2})},\\
\tilde{M}_{1}^{-}(y) &= M_{1}^{-}\big|_{y = -\frac{1}{2} - 0}e^{|\omega|(y + \frac{1}{2})}.
\end{aligned}
$$
Thereby, due to formulas \eqref{1.17} the components $L_{1}^{\pm}$ and $1 + M_{1}^{\pm}$ of the tension vector $\vec{H}$ of the magnetic field are defined in the domains $S_{1}^{\pm}$, and $L_{1}^{\pm} = 0$.

\section{Periodic perturbations. Linearized problem. Formulation of main results}
We will be looking for solutions of system \eqref{1.14} in the special form
\begin{equation}
\label{2.1}
\vec{U}(t,x,y) = \vec{\tilde{U}}(y)\exp\{\lambda t + i\omega x\},
\end{equation}
where $\lambda = \eta + i\xi$, $\xi,\omega \in R^{1}$,\\
$\vec{U} = (u, v, \alpha_{11}, \alpha_{12}, \alpha_{22}, \Omega, Z , L, M)^{T}$. We will below drop tildes. Then, it follows from \eqref{1.14} that
$$
\begin{aligned}
u^{'} &= \frac{\hat{\alpha}_{12}^{'} -i\omega\hat{\alpha}_{1}}{\hat{\alpha}_{2}}v + \frac{\lambda +i\omega\hat{u} + R_{44}}{\hat{\alpha}_{2}}\alpha_{12} + \frac{R_{43}}{\hat{\alpha}_{2}}\alpha_{11} + \frac{R_{45}}{\hat{\alpha}_{2}}\alpha_{22} + \frac{r_{12}}{\hat{\alpha}_{2}}Z,\\
v^{'} &= -i\omega u,\\
(\lambda &+ i\omega\hat{u} - \frac{2\hat{\alpha}_{12}R_{43}}{\hat{\alpha}_{2}} + R_{33})\alpha_{11} - 2\hat{\alpha}_{1}i\omega u - \left(\frac{2\hat{\alpha}_{12}(\hat{\alpha}_{12}^{'}-i\omega\hat{\alpha}_{1})}{\hat{\alpha}_{2}} -\hat{\alpha}_{11}^{'} \right)v -\\
&- \left(\frac{2\hat{\alpha}_{12}(\lambda + i\omega\hat{u} + R_{44})}{\hat{\alpha}_{2}} - R_{34} \right)\alpha_{12} - \left(\frac{2\hat{\alpha}_{12}R_{45}}{\hat{\alpha}_{2}} - R_{35} \right)\alpha_{22} -\\
&- \left(\frac{2\hat{\alpha}_{12}r_{12}}{\hat{\alpha}_{2}} - r_{11} \right)Z = 0,\\
\alpha_{12}^{'} &= \frac{\lambda + i\omega\hat{u}}{\hat{Z}}u + \frac{\hat{u}^{'}}{\hat{Z}}v - \frac{\hat{Z}^{'}}{\hat{Z}}\alpha_{12} + \frac{i\omega}{\hat{Z}}\Omega - i\omega\alpha_{11} + i\omega\alpha_{22} - \frac{i\omega\hat{\alpha}_{11} + \hat{\alpha}_{12}^{'}}{\hat{Z}}Z -\\
&- \frac{\hat{\alpha}_{12}}{\hat{Z}}Z^{'} + \frac{\sigma_{m}(1 + \hat{\lambda})}{\hat{Z}}(i\omega M - L^{'}) - \frac{\sigma_{m}\hat{L}^{'}}{\hat{Z}}M,\\
&(\lambda + i\omega\hat{u} + R_{55})\alpha_{22} - (2\hat{\alpha}_{12}i\omega - \hat{\alpha}_{22}^{'})v + 2\hat{\alpha}_{2}i\omega u + R_{53}\alpha_{11} +R_{54}\alpha_{12} = 0,\\
\Omega^{'} &= -(\lambda + i\omega\hat{u})v + i\omega\hat{Z}\alpha_{12} + (\hat{\alpha}_{12}i\omega + \hat{\alpha}_{22}^{'} + Gr)Z + \hat{\alpha}_{22}Z^{'} +\\
&+ \sigma_{m}\hat{L}(i\omega M - L^{'}) - \sigma_{m}\hat{L}^{'}L,\\
\frac{1}{Pr}Z^{''} &= (\frac{q_{2}}{Pr} - \frac{A_{r}}{Pr}(\hat{u}^{'}\hat{a}_{12} + \hat{Z}\frac{\hat{a}_{12}}{\hat{\alpha}_{2}}r_{12}) - \frac{A_{m}}{Pr}\sigma_{m}\hat{L}(1 + \hat{\lambda})\frac{r_{12}}{\hat{\alpha}_{2}})Z -\\
&- u(\frac{A_{r}}{Pr}\hat{Z}(\hat{a}_{11}i\omega - \hat{a}_{22})i\omega + \frac{A_{m}}{Pr}\sigma_{m}(\hat{L}^{2} - (1 + \hat{\lambda})^{2})i\omega) +\\
&+ v\bigg(\hat{Z}^{'} - \frac{A_{r}}{Pr}\hat{Z}(\hat{a}_{22}i\omega + \hat{a}_{12}\frac{\hat{\alpha}_{12} - i\omega\hat{\alpha}_{1}}{\hat{\alpha}_{2}}) - \frac{A_{m}}{Pr}\sigma_{m}\hat{L}(1 + \hat{\lambda})\times\\
&\times(i\omega + \frac{\hat{\alpha}_{12}^{'} - i\omega\hat{\alpha}_{1}}{\hat{\alpha}_{2}})\bigg) - \frac{A_{m}}{Pr}\sigma_{m}(1 + \hat{\lambda})\hat{u}^{'}L - \frac{A_{m}}{Pr}\sigma_{m}\hat{u}^{'}\hat{L}M -\\
&- \alpha_{12}\bigg(\frac{A_{r}}{Pr}\hat{Z}\hat{a}_{12} + \frac{A_{m}}{Pr}\sigma_{m}\hat{L}(1 + \hat{\lambda})\frac{\lambda + i\omega\hat{u} + R_{44}}{\hat{\alpha}_{2}} \bigg) -\\
&- \alpha_{11}\big(\frac{A_{r}}{Pr}\hat{Z}\hat{a}_{12} + \frac{A_{m}}{Pr}\sigma_{m}\hat{L}(1 + \hat{\lambda})\big)\frac{R_{43}}{\hat{\alpha}_{2}} - \alpha_{22}\big(\frac{A_{r}}{Pr}\hat{Z}\hat{a}_{12} + \frac{A_{m}}{Pr}\sigma_{m}\hat{L}(1 + \hat{\lambda})\big)\frac{R_{45}}{\hat{\alpha}_{2}},\\
L^{''} &= q_{3}L - \frac{i\omega\hat{L}}{b_{m}}u + v\bigg(\frac{\hat{L}^{'}}{b_{m}} - \frac{1 + \lambda}{b_{m}}\frac{(\hat{\alpha}_{12}^{'} - i\omega\hat{\alpha}_{1})}{\hat{\alpha}_{2}}\bigg) - \frac{1 + \hat{\lambda}}{b_{m}}\frac{\lambda + i\omega\hat{u} + R_{44}}{\hat{\alpha}_{2}}\alpha_{12} -
\end{aligned}
$$
\begin{equation}
\label{2.2}
\begin{aligned}
&- \frac{1 + \hat{\lambda}}{b_{m}}\frac{R_{43}}{\hat{\alpha}_{2}}\alpha_{11} - \frac{1 + \hat{\lambda}}{b_{m}}\frac{r_{12}}{\hat{\alpha}_{2}}Z - \frac{\hat{u}^{'}}{b_{m}}M - \frac{1 + \hat{\lambda}}{b_{m}}\frac{R_{45}}{\hat{\alpha}_{2}}\alpha_{22},\\
M^{''} &= q_{3}M + \frac{1 + \hat{\lambda}}{b_{m}}i\omega u - \frac{i\omega\hat{L}}{b_{m}}v,
\end{aligned}
\end{equation}

where $q_{2} = Pr(\lambda + i\omega\hat{u}) + \omega^{2}$, $q_{3} = \frac{\lambda + i\omega\hat{u} + b_{m}\omega^{2}}{b_{m}}$.

The following statements are true.

\begin{theorem}
If problem \eqref{1.14}, \eqref{1.15} has a solution in form \eqref{2.1} (the parameter $\omega$ is constant!), then we have the following asymptotic representation for $\lambda$:
\begin{multline}
\label{2.3}
\lambda_{k} = \left[\int_{-\frac{1}{2}}^{\frac{1}{2}}\frac{1}{\sqrt{\hat{Z}\hat{\alpha}_{2}}}d\xi \right]^{-1}\bigg(\int_{-\frac{1}{2}}^{\frac{1}{2}}-\frac{1}{2}\bigg[\sqrt{\frac{\hat{\alpha}_{2}}{\hat{Z}}}\bigg(\frac{i\omega\hat{u} + R_{44}}{\hat{\alpha}_{2}} + \frac{2\hat{\alpha}_{12}R_{43}}{\hat{\alpha}_{2}^{2}}\bigg) +\\
+i\omega\hat{u}\frac{1}{\sqrt{\hat{Z}\hat{\alpha}_{2}}} + \frac{\hat{\alpha}_{12}}{\hat{Z}}\times\\
\times (A_{r}\hat{Z}\hat{a}_{12} +A_{m}\sigma_{m}\hat{L}(1 + \hat{\lambda}))\sqrt{\frac{\hat{Z}}{\hat{\alpha}_{2}}} + \frac{\sigma_{m}(1 + \hat{\lambda})^{2}}{b_{m}\sqrt{\hat{Z}\hat{\alpha}_{2}}} \bigg]d\xi + k\pi i\bigg) + O(\frac{1}{k}), \quad k \to \infty,
\end{multline}
where we use $O$ as a big O notation.
\end{theorem}
From representation \eqref{2.3} we obtain a necessary condition for the asymptotic stability of the Poiseuille-type flow described in Sect.

\begin{theorem}
For the asymptotic stability of the Poiseuille-type flow it is necessary that the following inequality holds true:
\begin{multline}
\label{2.4}
\int_{-\frac{1}{2}}^{\frac{1}{2}}\bigg[\sqrt{\frac{\hat{\alpha}_{2}}{\hat{Z}}}\hat{\bar{\chi}}_{0}\left(\frac{1}{\hat{\alpha}_{2}}\big(W^{-1} + \frac{k + 2\beta}{3}(\hat{a}_{11} + \hat{a}_{22})\big) + \hat{\bar{\chi}}_{0}\frac{2\hat{\alpha}_{12}}{\hat{\alpha}_{2}^{2}}\hat{a}_{12}\frac{k + 2\beta}{3}\right) +\\
+ \frac{\hat{\alpha}_{12}}{\hat{Z}}\big(A_{r}\hat{Z}\hat{a}_{12} + A_{m}\sigma_{m}\hat{L}(1 + \hat{\lambda})\big)\sqrt{\frac{\hat{Z}}{\hat{\alpha}_{2}}} + \frac{\sigma_{m}(1 + \hat{\lambda})^{2}}{b_{m}\sqrt{\hat{Z}\hat{\alpha}_{2}}}\bigg]d\xi > 0.
\end{multline}
\end{theorem}

\section{Proof of theorems 1 and 2}
Due to the third and fifth equations of system \eqref{2.2} we can get representations for the components $\alpha_{11}$, $\alpha_{22}$:
\begin{equation}
\label{3.1}
\begin{aligned}
\alpha_{11} &= \frac{2\hat{\alpha}_{12}}{\hat{\alpha}_{2}}\alpha_{12} + \alpha_{11}^{*},\\
\alpha_{22} &= \alpha_{22}^{*},
\end{aligned}
\end{equation}
where the functions $\alpha_{11}^{*}$, $\alpha_{22}^{*}$ are expressed through $u,v,Z,\alpha_{12}$ with coefficients proportional to $\frac{1}{\lambda}$. We will further show that such values do not influence on the first term in the asymptotic representation for the eigenvalues $\lambda$ as $|\lambda| \to \infty$ and, hence, they can be omitted.

Using relations \eqref{3.1} and denoting
$$
\tilde{Y} = (\tilde{y}_{1}, \tilde{y}_{2}, \tilde{y}_{3}, \tilde{y}_{4}, \tilde{y}_{5}, \tilde{y}_{6}, \tilde{y}_{7}, \tilde{y}_{8}, \tilde{y}_{9}, \tilde{y}_{10})^{T} = (u,v,\alpha_{12},\Omega,Z,Z^{'},L,L^{'},M,M^{'})^{T},
$$
system \eqref{2.2} can be rewritten in the following form:
\begin{equation}
\label{3.2}
\begin{aligned}
\tilde{y}_{1}^{'} &= \frac{\hat{\alpha}_{12}^{'} -i\omega\hat{\alpha}_{1}}{\hat{\alpha}_{2}}\tilde{y}_{2} + \frac{\lambda +i\omega\hat{u} + R_{44} + \frac{2\hat{\alpha}_{12}R_{43}}{\hat{\alpha}_{2}}}{\hat{\alpha}_{2}}\tilde{y}_{3} + \frac{r_{12}}{\hat{\alpha}_{2}}\tilde{y}_{5},\\
\tilde{y}_{2}^{'} &= -i\omega\tilde{y}_{1},\\
\tilde{y}_{3}^{'} &= \frac{\lambda + i\omega\hat{u}}{\hat{Z}}\tilde{y}_{1} + \frac{\hat{u}^{'}}{\hat{Z}}\tilde{y}_{2} - (\frac{\hat{Z}^{'}}{\hat{Z}} + \frac{i\omega2\hat{\alpha}_{12}}{\hat{\alpha}_{2}})\tilde{y}_{3} + \frac{i\omega}{\hat{Z}}\tilde{y}_{4} - \frac{i\omega\hat{\alpha}_{11} + \hat{\alpha}_{12}^{'}}{\hat{Z}}\tilde{y}_{5} -\\
&- \frac{\hat{\alpha}_{12}}{\hat{Z}}\tilde{y}_{6} + \left(\frac{\sigma_{m}(1 + \hat{\lambda})}{\hat{Z}}i\omega - \frac{\sigma_{m}\hat{L}^{'}}{\hat{Z}}\right)\tilde{y}_{9} - \frac{\sigma_{m}(1 + \hat{\lambda})}{\hat{Z}}\tilde{y}_{8},\\
\tilde{y}_{4}^{'} &= -(\lambda + i\omega\hat{u})\tilde{y}_{2} + i\omega\hat{Z}\tilde{y}_{3} + (\hat{\alpha}_{12}i\omega + \hat{\alpha}_{22}^{'} + Gr)\tilde{y}_{5} - \sigma_{m}\hat{L}^{'}\tilde{y}_{7} - \sigma_{m}\hat{L}\tilde{y}_{8} +\\
&+ \sigma_{m}Li\omega\tilde{y}_{9} + \hat{\alpha}_{22}\tilde{y}_{6},\\
\tilde{y}_{5}^{'} &= \tilde{y}_{6},\\
\frac{1}{Pr}\tilde{y}_{6}^{'} &= (\lambda + i\omega\hat{u} + \frac{\omega^{2}}{Pr} - \frac{A_{r}}{Pr}(\hat{u}^{'}\hat{a}_{12} + \hat{Z}\frac{\hat{a}_{12}}{\hat{\alpha}_{2}}r_{12}) - \frac{A_{m}}{Pr}\sigma_{m}\hat{L}(1 + \hat{\lambda})\frac{r_{12}}{\hat{\alpha}_{2}})\tilde{y}_{5} -\\
&- \tilde{y}_{1}(\frac{A_{r}}{Pr}\hat{Z}(\hat{a}_{11} - \hat{a}_{22})i\omega + \frac{A_{m}}{Pr}\sigma_{m}(\hat{L}^{2} - (1 + \hat{\lambda})^{2})i\omega) +\\
&+ \tilde{y}_{2}\bigg(\hat{Z}^{'} - \frac{A_{r}}{Pr}\hat{Z}\hat{\alpha}_{12}(i\omega + \frac{\hat{\alpha}_{12}^{'} - i\omega\hat{\alpha}_{1}}{\hat{\alpha}_{2}}) - \frac{A_{m}}{Pr}\sigma_{m}\hat{L}(1 + \hat{\lambda})\times\\
&\times(i\omega + \frac{\hat{\alpha}_{12}^{'} - i\omega\hat{\alpha}_{1}}{\hat{\alpha}_{2}})\bigg) - \frac{A_{m}}{Pr}\sigma_{m}(1 + \hat{\lambda})\hat{u}^{'}\tilde{y}_{7} - \frac{A_{m}}{Pr}\sigma_{m}\hat{u}^{'}\hat{L}\tilde{y}_{9} -\\
&- \bigg(\frac{A_{r}}{Pr}\hat{Z}\hat{a}_{12} + \frac{A_{m}}{Pr}\sigma_{m}\hat{L}(1 + \hat{\lambda})\frac{\lambda + i\omega\hat{u} + R_{44} + \frac{2\hat{\alpha}_{12}R_{53}}{\hat{\alpha}_{2}}}{\hat{\alpha}_{2}} \bigg)\tilde{y}_{3},\\
\tilde{y}_{7}^{'} &= \tilde{y}_{8}\\
\tilde{y}_{8}^{'} &= (\frac{\lambda + i\omega\hat{u}}{b_{m}} + \omega^{2})\tilde{y}_{7} - \frac{i\omega\hat{L}}{b_{m}}\tilde{y}_{1} + \bigg(\frac{\hat{L}^{'}}{b_{m}} - \frac{1 + \lambda}{b_{m}}(\hat{\alpha}_{12} - i\omega\hat{\alpha}_{1})\bigg)\tilde{y}_{2} -\\
&- \frac{1 + \hat{\lambda}}{b_{m}}\frac{\lambda + i\omega\hat{u} + R_{44} + \frac{2\hat{\alpha}_{12}R_{43}}{\hat{\alpha}_{2}}}{\hat{\alpha}_{2}}\tilde{y}_{3} - \frac{1 + \hat{\lambda}}{b_{m}}{r_{12}}{\hat{\alpha}_{2}}\tilde{y}_{5} - \frac{\hat{u}^{'}}{b_{m}}\tilde{y}_{9},\\
\tilde{y}_{9}^{'} &= \tilde{y}_{10},\\
\tilde{y}_{10}^{'} &= \left(\frac{\lambda + i\omega\hat{u}}{b_{m}} + \omega^{2}\right)\tilde{y}_{9} + \frac{1 + \hat{\lambda}}{b_{m}}i\omega\tilde{y}_{1} - \frac{i\omega\hat{L}}{b_{m}}\tilde{y}_{2}.
\end{aligned}
\end{equation}
Let us rewrite system \eqref{3.2} in the more compact form
\begin{equation}
\label{3.3}
\tilde{Y}^{'} = (\lambda D + P)\tilde{Y},
\end{equation}
where the matrix $D$ is rather sparse:
$$
\begin{aligned}
d_{13} &= \frac{1}{\hat{\alpha}_{2}}, \, d_{31} = \frac{1}{\hat{Z}}, \, d_{42} = -1, \, d_{63} = -\frac{1}{\hat{\alpha}_{2}}(A_{r}\hat{Z}\hat{a}_{12} + A_{m}\sigma_{m}\hat{L}(1 + \hat{\lambda})),\\
d_{65} &= Pr, \, d_{83} = -\frac{1 + \hat{\lambda}}{b_{m}\hat{\alpha}_{2}}, \, d_{87} = \frac{1}{b_{m}}, \, d_{10,9} = \frac{1}{b_{m}},
\end{aligned}
$$
all other elements equal zero.\\
We can make the change
\begin{equation}
\label{3.4}
\tilde{Y} = TY,
\end{equation}
where the elements of the matrix $T$ read:
$$
\begin{aligned}
t_{11} &= -t_{12} = -\sqrt{\frac{\hat{Z}}{\hat{\alpha}_{2}}}, \, t_{71} = -t_{72} = (A_{r}\hat{Z}\hat{a}_{12} + A_{m}\sigma_{m}\hat{L}(1 + \hat{\lambda}))\sqrt{\frac{\hat{Z}}{\hat{\alpha}_{2}}},\\
t_{81} &= -t_{82} = \frac{1 + \hat{\lambda}}{b_{m}}\sqrt{\frac{\hat{Z}}{\hat{\alpha}_{2}}}, \, t_{58} = \frac{1}{Pr}, \, t_{7,10} = b_{m},\\
t_{31} &= t_{32} = t_{45} = t_{67} = t_{89} = t_{93} = t_{10,4} = -t_{26} = 1,
\end{aligned}
$$
and other elements are zero.\\
This enables one to write down the matrix $D$ in the upper Jordan form:
\begin{multline}
\label{3.5}
T^{-1}DT = W = diag\bigg\{-\frac{1}{\sqrt{\hat{Z}\hat{\alpha}_{2}}}, \frac{1}{\sqrt{\hat{Z}\hat{\alpha}_{2}}}, 0, 0, block\begin{pmatrix} 0 & 1\\ 0 & 0 \end{pmatrix}, block\begin{pmatrix} 0 & 1\\ 0 & 0 \end{pmatrix},\\
 block\begin{pmatrix} 0 & 1\\ 0 & 0 \end{pmatrix}\bigg\}.
\end{multline}
Then, system \eqref{3.3} is transformed in the following way:
\begin{equation}
\label{3.6}
Y^{'} = (\lambda W + C)Y,
\end{equation}
where $C = T^{-1}PT - T^{-1}T^{'}$.

Lets write only two elements of the matrix $C$ which will be used below:
\begin{equation}
\label{3.7}
\begin{aligned}
c_{11} &= -\frac{1}{2}\sqrt{\frac{\hat{\alpha}_{2}}{\hat{Z}}}\left(\frac{i\omega\hat{u} + R_{44} + \frac{2\hat{\alpha}_{12}R_{43}}{\hat{\alpha}_{2}}}{\hat{\alpha}_{2}} \right) + \frac{1}{2}\bigg(-i\omega\hat{u}\frac{1}{\sqrt{\hat{Z}\hat{\alpha}_{2}}} - \frac{\hat{Z}^{'}}{\hat{Z}} - \frac{i\omega\hat{\alpha}_{12}}{\hat{\alpha}_{2}} - \frac{\hat{\alpha}_{12}}{\hat{Z}}\times\\
&\times (A_{r}\hat{Z}\hat{a}_{12} + A_{m}\sigma_{m}\hat{L}(1 + \hat{\lambda}))\sqrt{\frac{\hat{Z}}{\hat{\alpha}_{2}}} - \frac{\sigma_{m}(1 + \hat{\lambda})^{2}}{b_{m}\sqrt{\hat{Z}\hat{\alpha}_{2}}}\bigg) - \frac{1}{2}\sqrt{\frac{\hat{\alpha}_{2}}{\hat{Z}}}\left(\sqrt{\frac{\hat{Z}}{\hat{\alpha}_{2}}}\right)^{'},\\
c_{22} &=\frac{1}{2}\sqrt{\frac{\hat{\alpha}_{2}}{\hat{Z}}}\left(\frac{i\omega\hat{u} + R_{44} + \frac{2\hat{\alpha}_{12}R_{43}}{\hat{\alpha}_{2}}}{\hat{\alpha}_{2}} \right) + \frac{1}{2}\bigg(-i\omega\hat{u}\frac{1}{\sqrt{\hat{Z}\hat{\alpha}_{2}}} - \frac{\hat{Z}^{'}}{\hat{Z}} - \frac{i\omega\hat{\alpha}_{12}}{\hat{\alpha}_{2}} - \frac{\hat{\alpha}_{12}}{\hat{Z}}\times\\
&\times (A_{r}\hat{Z}\hat{a}_{12} + A_{m}\sigma_{m}\hat{L}(1 + \hat{\lambda}))\sqrt{\frac{\hat{Z}}{\hat{\alpha}_{2}}} - \frac{\sigma_{m}(1 + \hat{\lambda})^{2}}{b_{m}\sqrt{\hat{Z}\hat{\alpha}_{2}}}\bigg) - \frac{1}{2}\sqrt{\frac{\hat{\alpha}_{2}}{\hat{Z}}}\left(\sqrt{\frac{\hat{Z}}{\hat{\alpha}_{2}}}\right)^{'}.
\end{aligned}
\end{equation}

Let us get an asymptotic representation for the fundamental matrix of system \eqref{3.6}.
For this purpose we split the matrix $C$ into blocks corresponding to representation \eqref{3.5} of the matrix $W$: the first diagonal block corresponds to the nonzero diagonal elements $-\frac{1}{\sqrt{\hat{Z}\hat{\alpha}_{2}}}$ and $\frac{1}{\sqrt{\hat{Z}\hat{\alpha}_{2}}}$ whereas the second diagonal block  corresponds, on the contrary, to zero elements. Then, system \eqref{3.6} can be written in the more convenient form
\begin{equation}
\label{3.8}
Y^{'} = \begin{pmatrix}Y_{I}\\ Y_{II} \end{pmatrix}^{'} = (\lambda W + C)\begin{pmatrix}Y_{I}\\ Y_{II} \end{pmatrix},
\end{equation}
where
\begin{equation}
\label{3.9}
C = \begin{pmatrix}
C_{I}^{I} & C_{II}^{I}\\
C_{I}^{II} & C_{II}^{II}
\end{pmatrix}.
\end{equation}
Using splitting \eqref{3.9} of the matrix $C$, we get the following system for the vector $Y_{II}$:
\begin{equation}
\label{3.10}
(Y_{II})^{'} = C_{II}^{II}Y_{II} + C_{I}^{II}Y_{I}.
\end{equation}
Assuming that the vector $Y_{I}$ is known, we can write down a system of fundamental solutions associated with the homogeneous system:
$$
Y_{II} = \sum_{i = 3}^{10}c_{i}Y_{II}^{i},
$$
where $c_{i}$ are arbitrary complex constants, $i = 3,\dots,10$, $Y^{i}_{II}\big|_{y = -\frac{1}{2}} = \begin{pmatrix} 0\\ \vdots \\ 0\\ 1 \\ 0 \\ \vdots \\ 0\end{pmatrix}$ is the $i$th component, and the general solution of system \eqref{3.10} is
\begin{equation}
\label{3.11}
Y_{II} = \sum_{i = 3}^{10} c_{i}Y_{II}^{i} + \int_{-\frac{1}{2}}^{y}Y(y)Y^{-1}(s)C_{I}^{II}Y_{I}ds,
\end{equation}
where $Y(y)$ is the fundamental matrix composed by the vectors $Y_{II}^{i}$.

Due to representation \eqref{3.11} the system for other component $Y_{I}$ can be written as
\begin{multline}
\label{3.12}
Y_{I}^{'} = \lambda\begin{pmatrix}
-\frac{1}{\sqrt{\hat{Z}\hat{\alpha}_{2}}} & 0\\
0 & \frac{1}{\sqrt{\hat{Z}\hat{\alpha}_{2}}}
\end{pmatrix}Y_{I} + C_{I}^{I}Y_{I} + C_{II}^{I}\left(\sum_{i = 3}^{10}c_{i}Y_{II}^{i}\right) + C_{II}^{I}\times\\
\times\int_{-\frac{1}{2}}^{y}Y(y)Y^{-1}(s)C_{I}^{II}Y_{I}(s)ds.
\end{multline}
Considering  $C_{II}^{I}(\sum\limits_{i = 3}^{10}c_{i}Y_{II}^{i})$ as a free term, we can find the fundamental matrix for system \eqref{3.12} in the following form:
\begin{equation}
\label{3.13}
\hat{Y} = \left(P_{0}(y) + \frac{1}{\lambda}P_{1}(y) + \frac{1}{\lambda^{2}}P_{2}(y) + \dots\right)\left(\delta_{ij}e^{\lambda\Gamma_{j}(y)}\right) + \frac{M_{0}}{\lambda^{2}} + \frac{M_{1}}{\lambda^{3}} + \dots, \, i,j = 1,2,
\end{equation}
where $\delta_{ij}$ is the Kronecker symbol, $\Gamma_{1}(y) = e^{-\int\limits_{-\frac{1}{2}}^{y}\frac{1}{\sqrt{\hat{Z}\hat{\alpha}_{2}}}d\xi}$,  $\Gamma_{2}(y) = e^{\int\limits_{-\frac{1}{2}}^{y}\frac{1}{\sqrt{\hat{Z}\hat{\alpha}_{2}}}d\xi}$.

\begin{remark}
The first term in \eqref{3.13} is the representation obtained by G. Birkhoff \cite{26} for the differential equation
$$
Y_{I}^{'} = \lambda\begin{pmatrix}
-\frac{1}{\sqrt{\hat{Z}\hat{\alpha}_{2}}} & 0\\
0 & \frac{1}{\sqrt{\hat{Z}\hat{\alpha}_{2}}}
\end{pmatrix}Y_{I} + C_{I}^{I}Y_{I}.
$$
\end{remark}
Substituting the matrix $\hat{Y}$ into equation \eqref{3.12}, we get the following correlation (the term $C_{II}^{I}(\sum\limits_{i = 3}^{10}c_{i}Y_{II}^{i})$ is omitted):
\begin{equation}
\label{3.14}
\begin{aligned}
&\left(P_{0}^{'}(y) + \frac{1}{\lambda}P_{1}^{'}(y) + \frac{1}{\lambda^{2}}P_{2}^{'}(y) + \dots\right)\left(\delta_{ij}e^{\lambda\Gamma_{j}(y)}\right)_{i,j = 1,2} +\\
&+ \lambda\left(P_{0}(y) + \frac{1}{\lambda}P_{1}(y) + \frac{1}{\lambda^{2}}P_{2}(y) + \dots\right)\Lambda\left(\delta_{ij}e^{\lambda\Gamma_{j}(y)}\right)_{i,j = 1,2} +\\
&+ \frac{M_{0}^{'}(y)}{\lambda^{2}} + \frac{M_{1}^{'}(y)}{\lambda^{3}} + \dots =\\
&= \lambda\Lambda\left(P_{0}(y) + \frac{1}{\lambda}P_{1}(y) + \frac{1}{\lambda^{2}}P_{2}(y) + \dots\right)\left(\delta_{ij}e^{\lambda\Gamma_{j}(y)}\right)_{i,j = 1,2} +\\
&+ \lambda\Lambda\left( \frac{M_{0}(y)}{\lambda^{2}} + \frac{M_{1}(y)}{\lambda^{3}} + \dots \right) +\\
&+ C_{I}^{I}\left(P_{0}(y) + \frac{1}{\lambda}P_{1}(y) + \frac{1}{\lambda^{2}}P_{2}(y) + \dots\right)\left(\delta_{ij}e^{\lambda\Gamma_{j}(y)}\right)_{i,j = 1,2} +\\
&+C_{I}^{I}\left( \frac{M_{0}(y)}{\lambda^{2}} + \frac{M_{1}(y)}{\lambda^{3}} + \dots \right) +\\
&+ C_{II}^{I}\int_{-\frac{1}{2}}^{y}Y(y)Y^{-1}(s)C_{I}^{II}\bigg[\left(P_{0}(s) + \frac{1}{\lambda}P_{1}(s) + \frac{1}{\lambda^{2}}P_{2}(s) + \dots\right)\times\\
&\times\left(\delta_{ij}e^{\lambda\Gamma_{j}(s)}\right)_{i,j = 1,2} + \frac{M_{0}(s)}{\lambda^{2}} + \frac{M_{1}(s)}{\lambda^{3}} + \dots\bigg]ds, \quad \Lambda = diag\{-\frac{1}{\sqrt{\hat{Z}\hat{\alpha}_{2}}}, \frac{1}{\sqrt{\hat{Z}\hat{\alpha}_{2}}}\}.
\end{aligned}
\end{equation}
Comparing the coefficients by the same powers of $\lambda$ which may contain the matrix $(\delta_{ij}e^{\lambda\Gamma_{j}(s)})_{i,j = 1,2}$ or be independent on it and integrating by parts a needed number of times, we, in particular, get
\begin{equation}
\label{3.15}
P_{0}\Lambda = \Lambda P_{0},
\end{equation}
\begin{equation}
\label{3.16}
P_{0}^{'} + P_{1}\Lambda = \Lambda P_{1} + C_{I}^{I}P_{0},
\end{equation}
\begin{equation}
\label{3.17}
P_{1}^{'} + P_{2}\Lambda = \Lambda P_{2} + C_{I}^{I}P_{1} + C_{II}^{I}C_{I}^{II}P_{0}\Lambda^{-1},
\end{equation}
\begin{equation}
\label{3.18}
\Lambda M_{0} = C_{II}^{I}YC_{I}^{II}(-\frac{1}{2}).
\end{equation}
Then, it follows from equality \eqref{3.15} that $P_{0}(y)$ is a diagonal matrix,
$$
P_{0}(y) = \begin{pmatrix}
p_{1}(y) & 0\\
0 & p_{2}(y)
\end{pmatrix},
$$
and equality \eqref{3.16} for the case of diagonal elements gives the Cauchy problems
\begin{equation}
\label{3.19}
p_{i}^{'} = c_{ii}^{I}p_{i}, \quad p_{i}(-\frac{1}{2}) = 1, \quad i = 1,2,
\end{equation}
where $c_{ii}^{I}$ are diagonal elements of the matrix $C_{I}^{I}$ (see formulas \eqref{3.7}).

Solving problems \eqref{3.19} gives us the functions $p_{i}(y)$: $p_{i}(y) = e^{\int\limits_{-\frac{1}{2}}^{y}c_{ii}^{I}(\xi)d\xi}$, $i = 1,2$ (the top index $I$ will be omitted below).

Then, equality \eqref{3.16} enables one to determine the non-diagonal elements of the matrix $P_{1}(y)$ which, in turn, enables one to find the diagonal elements from equality \eqref{3.17}. By finite induction we can find all the matrices $P_{i}(y)$, $i = 2,\dots$. Equalities analogous to \eqref{3.18} enables one to determine the matrices $M_{i}$, $i = 0,1,\dots$

Due to the method of variation of constants the presence of the free term $C_{II}^{I}(\sum\limits_{i = 3}^{10}c_{i}Y_{II}^{i})$  results in representation \eqref{3.13} for the additional terms containing as multipliers the powers of $\frac{1}{\lambda}$: $\frac{1}{\lambda^{k}}$, $k = 1,2,\dots$. It becomes then clear that such terms do not influence on the main term in the asymptotic representation of the spectrum. This means that we can use only the main term in representation \eqref{3.13}:
\begin{equation}
\label{3.20}
Y = P_{0}(y)(\delta_{ij}e^{\lambda\Gamma_{j}(y)})_{i,j = 1,2}.
\end{equation}
Recalling the fundamental matrix of equation \eqref{3.8} and taking into account \eqref{3.20}, we get the main term in the asymptotic representation of the fundamental matrix $W_{Y}$ of system \eqref{3.8}:
\begin{equation}
\label{3.21}
W_{Y} = \begin{pmatrix}
e^{-\lambda\int\limits_{-\frac{1}{2}}^{y}\frac{1}{\sqrt{\hat{Z}\hat{\alpha}_{2}}}d\xi}p_{1}(y) & 0 & 0 & \dots & 0\\
0 & e^{\lambda\int\limits_{-\frac{1}{2}}^{y}\frac{1}{\sqrt{\hat{Z}\hat{\alpha}_{2}}}d\xi}p_{2}(y) & 0 & \dots & 0\\
0 & 0 & y_{3}^{3} & \dots & y_{10}^{3}\\
\dots & \dots & \dots & \dots & \dots\\
0 & 0 & y_{3}^{10} & \dots & y_{10}^{10}\\
\end{pmatrix},
\end{equation}
where $y_{i}^{j}$, $i,j = 3,\dots,10$ are components of the fundamental system for equation \eqref{3.10} composed by the columns of the matrix $Y(y)$.

\begin{remark}
In this work we do not give a justification of the fact that representation \eqref{3.13} is really asymptotic series as well as we do not justify the described representation of the fundamental matrix $W_{Y}$ (more precisely, its ``full variant''). This is a subject for future research. We only note that this fact was established by G. Birkhoff \cite{26} for  equation \eqref{3.14} (when there are no terms with coefficients $M_{0}$, $M_{1}$, $\dots$ in the right-hand side of equality \eqref{3.13}, i.e., the part of the series corresponding to the integral term in  equation \eqref{3.12}), and in each half-plane $Re \lambda > 0$ and $Re \lambda < 0$ the asymptotic series are different from each other \cite{27}.
\end{remark}

By the way, considering the integral term as a free term and using the method of variation of constants is another way of finding the matrices $M_{0}$, $M_{1}$, $\dots$ (the idea of getting fundamental matrices by the method of variation of constants is described in \cite{28}).

Recalling the boundary conditions for $y = \pm\frac{1}{2}$, we can note that relations \eqref{1.15} after change \eqref{3.4} of the variable $\tilde{Y}$ are transformed in the following way:
\begin{equation}
\label{3.22}
y_{1} = y_{2}, \, y_{6} = 0, \, y_{8} = 0, \, y_{10} = 0, \, y_{4} = 0, \quad \mbox{при } y = \pm\frac{1}{2}, \, t > 0.
\end{equation}
Or, taking into account representation \eqref{3.21}, they can be written as the equality
\begin{equation}
\label{3.23}
det\begin{pmatrix}
L\\
LW_{Y}(\frac{1}{2})
\end{pmatrix} = 0,
\end{equation}
where
\setcounter{MaxMatrixCols}{20}
$$
L = \begin{pmatrix}
1 & -1 & 0 & 0 & 0 & 0 & 0 & 0 & 0 & 0 & 0\\
0 & 0 & 0 & 0 & 0 & 1 & 0 & 0 & 0 & 0 & 0\\
0 & 0 & 0 & 0 & 0 & 0 & 0 & 1 & 0 & 0 & 0\\
0 & 0 & 0 & 0 & 0 & 0 & 0 & 0 & 0 & 0 & 1\\
0 & 0 & 0 & 1 & 0 & 0 & 0 & 0 & 0 & 0 & 0
\end{pmatrix}.
$$
After elementary transforms of the determinant and the use of the Laplace theorem about the representation of the determinant as multiplications of its minors we see that \eqref{3.23} is equivalent to the equality
\begin{equation}
\label{3.24}
\begin{aligned}
&e^{\lambda\int\limits_{-\frac{1}{2}}^{\frac{1}{2}}\frac{1}{\sqrt{\hat{Z}\hat{\alpha}_{2}}}d\xi}p_{2}(\frac{1}{2}) - e^{-\lambda\int\limits_{-\frac{1}{2}}^{\frac{1}{2}}\frac{1}{\sqrt{\hat{Z}\hat{\alpha}_{2}}}d\xi}p_{1}(\frac{1}{2}) = 0, \, \mbox{или}\\
&e^{\lambda\int\limits_{-\frac{1}{2}}^{\frac{1}{2}}\frac{1}{\sqrt{\hat{Z}\hat{\alpha}_{2}}}d\xi}e^{\int\limits_{-\frac{1}{2}}^{\frac{1}{2}}c_{22}(\xi)d\xi} - e^{-\lambda\int\limits_{-\frac{1}{2}}^{\frac{1}{2}}\frac{1}{\sqrt{\hat{Z}\hat{\alpha}_{2}}}d\xi}e^{\int\limits_{-\frac{1}{2}}^{\frac{1}{2}}c_{11}(\xi)d\xi} = 0.
\end{aligned}
\end{equation}
Recalling formulas \eqref{3.7}, we get the spectrum representation
\begin{multline}
\label{3.25}
\lambda_{k} = \left[\int_{-\frac{1}{2}}^{\frac{1}{2}}\frac{1}{\sqrt{\hat{Z}\hat{\alpha}_{2}}}d\xi \right]^{-1}\bigg(\int_{-\frac{1}{2}}^{\frac{1}{2}}-\frac{1}{2}\bigg[\sqrt{\frac{\hat{\alpha}_{2}}{\hat{Z}}}\bigg(\frac{i\omega\hat{u} + R_{44}}{\hat{\alpha}_{2}} + \frac{2\hat{\alpha}_{12}R_{43}}{\hat{\alpha}_{2}^{2}}\bigg) +\\
+i\omega\hat{u}\frac{1}{\sqrt{\hat{Z}\hat{\alpha}_{2}}} + \frac{\hat{\alpha}_{12}}{\hat{Z}}\times\\
\times (A_{r}\hat{Z}\hat{a}_{12} +A_{m}\sigma_{m}\hat{L}(1 + \hat{\lambda}))\sqrt{\frac{\hat{Z}}{\hat{\alpha}_{2}}} + \frac{\sigma_{m}(1 + \hat{\lambda})^{2}}{b_{m}\sqrt{\hat{Z}\hat{\alpha}_{2}}} \bigg]d\xi + k\pi i\bigg) + O(\frac{1}{k}), \quad k \to \infty,
\end{multline}
where the symbol $O$ denotes a big O.\\
The proof of Theorem 1 is thus complete.
\begin{remark}
Using representation \eqref{3.13}, we can get an asymptotic representation of $\lambda_{k}$ with an arbitrary order of accuracy defined by the powers $\frac{1}{k}$ (see also \cite{26}).
\end{remark}

Now, as a consequence of representation \eqref{3.25}, we get the following result. If the  Poiseuille-type flow described in Sect. 1 is asymptotically stable, then the following inequality necessarily  hold:
\begin{multline*}
Re\lambda_{k} = \int_{-\frac{1}{2}}^{\frac{1}{2}}\bigg[\sqrt{\frac{\hat{\alpha}_{2}}{\hat{Z}}}(\frac{R_{44}}{\hat{\alpha}_{2}} + \frac{2\hat{\alpha}_{12}R_{43}}{\hat{\alpha}_{2}^{2}}) +\\
+ \frac{\hat{\alpha}_{12}}{\hat{Z}}\big(A_{r}\hat{Z}\hat{a}_{12} + A_{m}\sigma_{m}\hat{L}(1 + \hat{\lambda})\big)\sqrt{\frac{\hat{Z}}{\hat{\alpha}_{2}}} + \frac{\sigma_{m}(1 + \hat{\lambda})^{2}}{b_{m}\sqrt{\hat{Z}\hat{\alpha}_{2}}}\bigg]d\xi > 0,
\end{multline*}
or taking into account formula \eqref{1.16},
\begin{multline}
\label{3.27}
Re\lambda_{k} = \int_{-\frac{1}{2}}^{\frac{1}{2}}\bigg[\sqrt{\frac{\hat{\alpha}_{2}}{\hat{Z}}}\hat{\bar{\chi}}_{0}\left(\frac{1}{\hat{\alpha}_{2}}\big(W^{-1} + \frac{k + 2\beta}{3}(\hat{a}_{11} + \hat{a}_{22})\big) + \hat{\bar{\chi}}_{0}\frac{2\hat{\alpha}_{12}}{\hat{\alpha}_{2}^{2}}\hat{a}_{12}\frac{k + 2\beta}{3}\right) +\\
+ \frac{\hat{\alpha}_{12}}{\hat{Z}}\big(A_{r}\hat{Z}\hat{a}_{12} + A_{m}\sigma_{m}\hat{L}(1 + \hat{\lambda})\big)\sqrt{\frac{\hat{Z}}{\hat{\alpha}_{2}}} + \frac{\sigma_{m}(1 + \hat{\lambda})^{2}}{b_{m}\sqrt{\hat{Z}\hat{\alpha}_{2}}}\bigg]d\xi > 0.
\end{multline}
The proof of Theorem 2 is complete.

The authors are grateful to A.V. Yegitov for his help in the preparation of the manuscript of the paper.\\
This work is supported by the Russian Foundation for Basic Research (the grant numbers 17-01-00791а and 19-01-00261а).

\end{document}